\documentclass[aps,prl,showpacs,showkeywords,twocolumn,superscriptaddress]{revtex4-1}

\usepackage{graphicx, graphics}% Include figure files
\usepackage{mathcomp}
\usepackage{psfrag}
\usepackage{amsmath,amssymb,lmodern}
\usepackage{bm}% bold math
\usepackage[below]{placeins}
\usepackage{array}
\usepackage[usenames]{color}
\usepackage{textcomp}
\begin{document}

\title{Lattice supersolid phase of strongly correlated bosons \\ in an optical cavity}
\author{Yongqiang Li}
\affiliation{Institut f\"ur Theoretische Physik, Goethe-Universit\"at, 60438 Frankfurt/Main, Germany}
\affiliation{Department of Physics, National University of Defense Technology, Changsha 410073, P. R. China}
\author{Liang He}
\affiliation{Institut f\"ur Theoretische Physik, Goethe-Universit\"at, 60438 Frankfurt/Main, Germany}
\author{Walter Hofstetter}
\affiliation{Institut f\"ur Theoretische Physik, Goethe-Universit\"at, 60438 Frankfurt/Main, Germany}
\date{\today}

\begin{abstract}
We numerically simulate strongly correlated ultracold bosons coupled to a high-finesse cavity field, pumped by a laser beam in the transverse direction. Assuming a weak $\it classical$ optical lattice added in the cavity direction, we model this system by a generalized Bose-Hubbard model, which is solved by means of bosonic dynamical mean-field theory. The complete phase diagram is established, which contains two novel self-organized
quantum phases, lattice supersolid and checkerboard solid, in addition to conventional
phases such as superfluid and Mott insulator. At finite but low temperature, thermal fluctuations are found to enhance the buildup of
the self-organized phases. We demonstrate that cavity-mediated long-range interactions can give rise to stable lattice
supersolid and checkerboard solid phases even in the regime of strong $s$-wave scattering.
In the presence of a harmonic trap, we discuss coexistence of these self-organized phases, as relevant to experiments.
\end{abstract}


\maketitle
Experimental realizations of atomic many-body systems coupled to a high-finesse cavity have recently attracted a large amount of attention
\cite{M. Lewenstein_2008}. In particular, the self-organized phase of
atoms induced by coherent scattering between pump laser and cavity mode has been predicted theoretically~\cite{H.
Ritsch_2002}, and confirmed experimentally for
laser-cooled atoms in a transversally pumped cavity \cite{A.T.
Black_2003}. However, only recently it has become possible to combine a high-finesse cavity with an ultracold
quantum gas in the strong-coupling regime and to experimentally investigate properties
of a Bose-Einstein condensate (BEC) in an optical cavity~\cite{J. Reichel_2007_1, J. Reichel_2007_2, P.W. Courteille_2007, T. Esslinger_2006_2009_1}. A phase transition from a normal to self-organized phase in an open system has been realized~\cite{T. Esslinger_2010}, and a lifetime up to 10ms of the self-organized phase has been achieved which indicates a steady state. Up to now, these experiments have, however, focused only on weakly interacting condensates. On the theory side, there is a lack of quantitative predictions for strongly correlated bosons coupled to an optical cavity, even though an extended Bose-Hubbard model has been derived \cite{C. Maschler_2005, J. Larson_2008} which describes the ultracold gas
trapped in a periodic optical potential generated by the high-finesse
cavity. Recently, theoretical studies of the BEC-cavity
system have predicted that the ground state can be Mott insulating with finite photon excitations of the cavity mode \cite{B. D. Simons_2009, G.
Morigi_2010}. However, the robustness of this self-organized phase
against strong contact interactions, finite temperature and the inhomogeneity induced by an external trap remains an important open issue.

To bridge this gap, here we numerically investigate the buildup of self-organized phases in ultracold bosonic gases coupled to a single-mode
cavity field, pumped by a laser beam in the transverse direction. This setup is similar to a two-dimensional (2D) classical optical lattice but with a quantized field in the cavity direction. Since the cavity field mediates long-range interactions between atoms~\cite{H. Ritsch_2004, P. Domokos_2007, T. Esslinger_2012}, we investigate the system by means of \emph{real-space} bosonic dynamical mean-field theory (RBDMFT) which captures both strong correlations and spatial inhomogeneity as well as arbitrary long-range order in a unified framework \cite{RBDMFT}.

Motivated by the recent experiment \cite{T. Esslinger_2010}, we
consider a system of ultracold $^{87}$Rb atoms with natural $s$-wave scattering length ${\tilde a}_s=5.77$
nm and atomic transition wavelength $\lambda=780.2$ nm, which is
driven by a linearly polarized standing-wave laser with a red-detuned
wave length $\lambda_p=784.5$ nm in the direction perpendicular to
the cavity axis. The setup of our simulation consists of the optical cavity in the $x$ direction, driven by a pump laser in the $z$ direction, and a strong confinement freezing the motional degree of
freedom of the atoms in the third direction \cite{third_direction}. We choose the cavity decay rate as $\kappa=300\omega_R$ which
is close to the experimental value of $\kappa = 2\pi \times
1.3$ MHz \cite{T. Esslinger_2010}, where $\omega_R$ is the frequency
corresponding to the recoil energy, $E_R = \hbar \omega_R =
h^2/(2m\lambda^2_p)$($\approx 2\pi\times3.8$ kHz). We choose the light shift as
$U_0=g^2_0/\Delta_a = -0.1\omega_R$, which leads to an atom-cavity coupling
strength $g_0$ two orders of magnitude larger than the cavity decay rate $\kappa$
and thus implies that the system is in the strong-coupling regime of cavity QED \cite{T. Esslinger_2006_2009_1}, where $\Delta_a$ denotes the atom-pump detuning.

This system can be described by an extended Bose-Hubbard model
\cite{C. Maschler_2005, J. Larson_2008}, where, for generality, a weak classical optical lattice is added in the cavity direction. We further assume the cavity mode to be in a coherent state to simplify the atom-cavity coupling, which is in good agreement with experimental results~\cite{T.
Esslinger_2010}. Within this approximation, the cavity
mode is described by a complex amplitude $\alpha$, and the
parameters of the extended Bose-Hubbard model only depend on the
average photon numbers. We thus finally obtain the lowest-band effective
Hamiltonian employed in the following calculations:
\begin{eqnarray}\label{Hamil}
\hat{H} &=& - \sum_{\langle i,j \rangle}\tilde{J}_{x(z)}\hat{b}^\dagger_i\hat{b}_j
           +\frac12U\sum_{i}\hat{b}^\dagger_i\hat{b}^\dagger_i\hat{b}_i\hat{b}_i \nonumber\\
        && + 2Re[\alpha] \eta_{\rm eff}J^\prime_{0}\sum_{i}(-1)^{i}\hat{b}^\dagger_i\hat{b}_i \nonumber\\
        && + \sum_{i}(V_i - \tilde{\mu} )\hat{b}^\dagger_i \hat{b}_i
\end{eqnarray}
where $\hat{b}^{\dagger}_{i}$ ($\hat{b}_{i}$) denotes the bosonic creation (annihilation) operator for a Wannier state at site $i$. Here $\tilde{J}_x$ ($\tilde{J}_z$) is the effective nearest-neighbor hopping amplitude in the $x$ ($z$) direction, with the hopping in the $x$ direction determined by the cavity mode, $\tilde{\mu}$ is the effective chemical potential, $V_i=V_{\rm trap}\, i^2$ with the strength $V_{\rm trap}$ of the external harmonic trap, and $U=4\pi a_s \hbar^2/m$ is the Hubbard interaction strength. The cavity mode amplitude $\alpha = {\eta_{\rm eff}J^\prime_{0} \sum_{i}(-1)^{i} \langle \hat{b}^\dagger_i\hat{b}_i \rangle} / {(\Delta^\prime_c + i\kappa)}$~\cite {P. Domokos_2008} with $\Delta^\prime_c=\Delta_c - U_0 (J^c_{0}\sum_{i}\langle \hat{b}^\dagger_i \hat{b}_i\rangle + J^c_1 \sum_{\langle i,j \rangle}\langle \hat{b}^\dagger_i\hat{b}_j \rangle)$ is determined self-consistently by the density distribution of the atoms. $J^c_0$, $J^\prime_{0}$ and $J^c_1$ denote the on-site single-particle matrix elements of the potential generated by the cavity mode, by scattering between pump laser and cavity mode via single atoms, and the first-order tunneling matrix element between nearest-neighbor sites of the cavity mode standing wave, respectively. $\eta_{\rm eff}=-\sqrt{|V_pU_0|}$ denotes the effective pump strength into the cavity through atomic scattering, $\Delta_c$ the cavity-pump detuning, and $V_p$ ($V_p=V_z$) the depth of the standing-wave potential created by the pump laser in the $z$ direction. The hopping amplitudes in the $x$ and $z$ direction for nearest neighbors are given by $\tilde
J_{x,z}/E_R=(4/\sqrt{\pi})(V_{x,z}/E_R)^{(3/4)}
\exp(-2\sqrt{V_{x,z}/E_R})$ and the Hubbard interaction parameter by $U/{E_R} =
4\sqrt{2\pi} (a_s/\lambda_p) (V_xV_zV_y/E^3_R)^{(1/4)}$ \cite{W.
Werger_2003}, where $V_{x}$~($V_y,V_z$) is the optical lattice depth in the $x$ ($y$, $z$) direction and $V_x$ is self-consistently determined by the cavity mode. For the onsite coupling matrix elements we use a Gaussian approximation of the Wannier states. To ensure that the tight-binding approximation is valid, we assume an external optical lattice in the cavity direction with a depth of $V_{\rm ext}=5E_R$.

The main challenge now is to determine the steady state of the BEC-cavity system described by the Hamiltonian (\ref{Hamil}) in the
Wannier basis. Here we apply real-space bosonic dynamical mean-field theory (RBDMFT) \cite{RBDMFT} which provides a nonperturbative description of the many-body system both in three and two spatial dimensions (considered here)~\cite{BDMFT_2009}. RBDMFT, which is capable of including the inhomogeneity of a trapped system as well as strong correlations between the atoms, assumes the self-energy to be local but site dependent. In our calculations, we choose the recoil energy $E_R$ ($\omega_R$) as the unit of energy, and set $\hbar=1$.

We first investigate the robustness of the supersolid phase against interactions for experimentally relevant parameters.
The supersolid is characterized by coexistence of the staggered order parameter $\Phi = {\langle \sum_{i}(-1)^{i}b^\dagger_ib_i\rangle}/{ \langle \sum_{i}b^\dagger_i b_i \rangle}$ and superfluid order $\phi = \langle b \rangle$. There are two possible signs of $\Phi$; i.e., the majority of the atoms occupy even sites for $\Phi>0$ or odd sites for $\Phi<0$ \cite{T. Esslinger_2006_2009_1}. Intuitively, if the pump laser is strong enough to stabilize a larger atom density at the even sites, and at the same time we choose a negative shifted cavity detuning $\Delta^\prime_c<0$, this implies that the coherent scattering between the pump laser and the cavity mode generates a potential with minima at the even sites, as indicated by the staggered term of Eq. (\ref{Hamil}). As a result, the corresponding potential will attract more atoms toward even sites and the system self-organizes into a steady state. In the following, we will confirm this heuristic argument via numerical simulations based on RBDMFT.
\begin{figure}[h]
\vspace{-12mm}
\includegraphics*[width=1.15\linewidth]{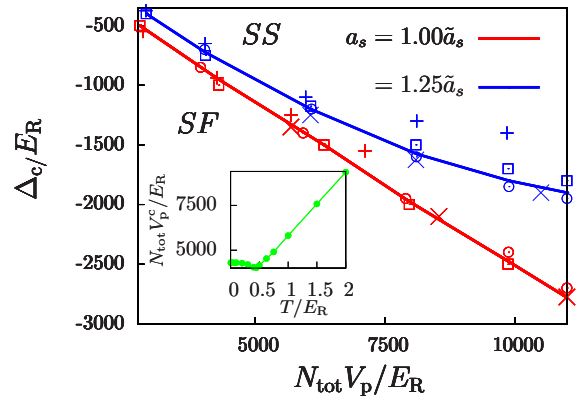}
\vspace{-7mm}
\caption{\!\!\!(Color online) Zero-temperature phase diagram at filling $n=1.98$ in terms of cavity-pump detuning $\Delta_{\rm c}$ and {\em rescaled} pump-laser power $N_{\rm tot} V_p$ for two different scattering lengths $a_s=\tilde{a}_s$ [red (lower) line], and $1.25\tilde{a}_s$ [blue (upper) line] (${\tilde a}_s = 5.77$ nm). There are two phases in the diagram: superfluid (SF) and supersolid (SS). Different markers correspond to different sizes of the system in our calculations [$N_{\rm lat} = 12\times12 \,(+)$, $N_{\rm lat} = 16\times16 \,(\boxdot)$, $N_{\rm lat} = 20\times20 \,(\odot)$, and $N_{\rm lat} = 24\times24\,(\times)$]. The cavity decay rate is set to $\kappa=300E_R$ and the light shift is $U_0=-0.1E_R$. Inset: Rescaled critical strength $N_{\rm tot} V^{\rm c}_{\rm p}$ of the standing-wave pump laser vs. temperature at fixed cavity detuning $\Delta_c=-1000E_R$ obtained from calculations on a $16\times16$ lattice.
}
\label{phase_diagram}
\end{figure}

Figure~\ref{phase_diagram} shows a zero-temperature phase diagram at filling $n=1.98$ in terms of cavity-pump detuning $\Delta_{\rm c}$ and {\em rescaled} pump-laser power $N_{\rm tot} V_p$ for two different scattering lengths $a_s=\tilde{a}_s, \, 1.25\tilde{a}_s$ (${\tilde a}_s = 5.77$ nm). One reason for using the rescaled pump-laser power $N_{\rm tot} V_p$ in the phase diagram is that the system has a physical finite size effect, meaning that in general, at a given filling, the phase boundary in terms of $\Delta_c$ and $V_p$ depends on the system size. The physical origin of this effect is that the pump laser globally couples to all atoms in the cavity, thus the strength of the coherently scattered light field in the cavity direction is proportional to the atom number, which in turn shifts the transition boundary. This motivates us to rescale $V_p$ by the total atom number. For a sufficiently large total particle number (according to our calculations $N_{\rm tot}\gtrsim500$), we indeed observe a nearly universal phase boundary regardless of the system size. For a weak pump laser, the system is superfluid with homogeneous density distribution and $\Phi=0$. In this case, the mean photon number in the cavity is zero. On the other hand, if the pump laser is strong enough, more photons are scattered into the cavity mode and the atoms organize themselves into a checkerboard pattern with $|\Phi|>0$. Our simulations thus clearly confirm the existence of the supersolid phase for single-component Bose gases in the cavity in the presence of strong on-site interactions. In Fig.~\ref{phase_diagram} we observe that the trend of the phase boundary from superfluid to supersolid as a function of $V_p$ at fixed scattering length $a_s$ is consistent with experiment~\cite{T. Esslinger_2010}. The phase boundary is considerably shifted upwards for larger scattering length, which indicates that more pump laser power is needed to drive the system into the self-organized phase. We also observe that onsite interactions have a more pronounced effect on the buildup of the supersolid phase for a stronger pump laser field. Generally, there exists also an unstable state for positive shifted cavity detuning $\Delta^\prime_c > 0$ \cite{H. Ritsch_phase, T. Esslinger_2010}, which is beyond the scope of this work.

We also investigate the effect of finite temperature on the critical pump strength, as shown in Fig.~\ref{phase_diagram}(a). We observe a minimum of $V^c_p$ at low but finite temperature, since thermal fluctuations excite the atoms from the ground state and thus reduce the energy gap between the homogeneous and the self-organized state. As a result, less power of the pump laser is needed to stabilize the supersolid. On the other hand, at high temperature, thermal fluctuations tend to smear out the self-organized density pattern, and as a result, more power is needed to stabilize it. Interestingly, the maximum of checker-board order occurs when the superfluid order vanishes. A similar effect in a different model has been observed in Ref.~\cite{Boninsegni}. Note that the long-range order $\phi \neq 0$ at $T > 0$ in two dimensions is a mean-field artifact in the thermodynamical limit, while in reality, the system exhibits a Kosterlitz-Thouless transition~\cite{finite-2D}.

%Another issue needed to be addressed is finite size effect for the homogeneous system. In general, at the same filling, the phase boundary in terms of $\Delta_c$ and $V_p$ depends on the system size. The physical origin of this effect is that the pump laser globally couples to all the atoms in the cavity, thus the strength of the coherently scattered light field in the cavity direction yields an enhancement propotional to the atom number which in turn shifts the transition boundary. This motivates us to rescale $V_p$ by multiplying it with the total atom number. As shown in Fig. 1(b), for large total particle number, we do observe an universal phase boundary regardless of the system size, which can acts as a useful guidence for relevant experiments addressing the transition between these two phases.
%As shown in Fig.~\ref{phase_diagram}(b), the critical pump strength $V^{\rm c}_{\rm p}$ decreases for a larger system but $N_{\rm tot}V^{\rm c}_{\rm p}$ is a constant for different total particle numbers, since the superfluid-to-supersolid transition is due to collective effects of all atoms in the cavity. It indicates that the scaled phase boundary at a fixed filling is universal in the superfluid regime studied here, whose signature can be directly used for prediction of phase boundary for ongoing experiments.

\begin{figure}[h]
\centering
\vspace{-8mm}
\includegraphics*[width=3.6in]{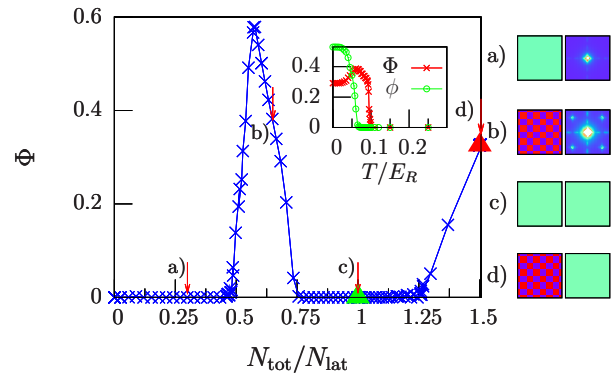}
\vspace{-35pt}
\caption{(Color online) Properties of the self-organized phases of
strongly interacting bosons on a square ($N_{\rm lat}=16\times16$) lattice. The blue curve corresponds to the filling dependence of checkerboard order $\Phi$ at zero temperature; the green (lower) and the red (upper) triangles denote  where the system is in the Mott insulator and checkerboard solid, respectively. (a)--(d): Density distribution of superfluid, supersolid, Mott insulator, and
checkerboard solid, respectively, in real space (left) and in quasi-momentum space (right), corresponding to the densities marked by the red arrows in the main figure. Other parameters are $\Delta_c=-500E_R$, $\kappa=300E_R$, $U_0=-0.1E_R$, and $V_p=15E_R$.
Inset: Melting of the supersolid phase with increasing temperature at fixed filling $N_{\rm tot}/N_{\rm lat}=0.68$, where the green (circle symbols) and red (star symbols) curves indicate the temperature dependence of the superfluid order $\phi$ and checkerboard order $\Phi$.}\label{phase_transition}
\end{figure}
From the previous discussion, we conclude that on-site interactions strongly shift the phase boundary between superfluid and supersolid. The sensitivity to on-site interactions has been also observed experimentally in Ref.~\cite{T. Esslinger_2010}.
We now investigate this effect in detail at different fillings on a square ($N_{\rm lat}=16\times16$) lattice. We choose a cavity detuning $\Delta_c = -500\omega_R$, a scattering length of $2.5{\tilde a}_s$, and a lattice depth $V_p=15E_R$ of the standing-wave pump laser, motivated by the recent experiment \cite{T. Esslinger_2010}. Figure~\ref{phase_transition} displays the resulting checkerboard order $\Phi$ (blue line) as a function of filling, where four possible phases of the BEC-cavity system are observed. Panels (a)--(d) in Fig.~\ref{phase_transition} show the density distribution in real space (left) and in quasimomentum space (right): (a) superfluid phase ($\phi \neq 0$ and $\Phi = 0$) with off-diagonal long-range order (phase coherence), (b) supersolid ($\phi \neq 0$ and $\Phi \neq 0$) with coexisting diagonal long-range order (periodic density modulation) and phase coherence, (c) Mott insulator ($\phi = 0$ and $\Phi = 0$) with zero mean-photon number in the cavity mode, and (d) checkerboard solid ($\phi = 0$ and $\Phi \neq 0$) with diagonal long-range order and finite mean-photon number in the cavity mode. Let us now discuss the underlying mechanism for the buildup of the self-organized phases. The excitation of the cavity mode is a collective effect due to all the atoms in the cavity and depends on the total particle number; i.e., the more atoms are in the cavity, the more photons will be coherently scattered into the cavity mode, and the easier the checkerboard pattern of the density distribution can be formed. In the absence of induced long-range interactions, there are two possible phases for strongly interacting bosonic gases in an optical lattice: superfluid and Mott insulator. The low-lying excitations of the superfluid phase are gapless sound modes which can be easily excited~\cite{W. Hofstetter_2010}, while the lowest excitations of the Mott insulator are gapped particle-hole pairs with an energy gap of order $U$~\cite{I. bloch_2002}. These different excitation properties, which can be detected via Bragg spectroscopy~\cite{W. Hofstetter_2010, T. Esslinger_2012}, strongly influence the buildup of the self-organized phases. As can be seen from the blue curve in Fig.~\ref{phase_transition}, the order parameter $\Phi$ becomes finite with increasing total particle number, and decreases to zero again in the vicinity of the Mott insulator. With further increase of the filling $n>1$, the checkerboard supersolid phase appears again. Interestingly, there is also a
checkerboard $solid$ phase emerging at $n=1.5$, since for larger particle number more photons are scattered into the cavity mode, and the resulting standing wave in the cavity direction suppresses tunneling of atoms and therefore superfluidity. Interestingly, we observe a maximum of the order parameter $\Phi$ at finite temperature due to the competition between superfluid and checkerboard order. All four phases can be detected experimentally by combining time-of-flight measurements and the detection of photons leaking from the cavity~\cite{T. Esslinger_2010}.

We have so far studied the homogeneous case, but in
real experiments the external trap induces inhomogeneity and a resulting coexistence of superfluid, Mott insulator, supersolid, and checkerboard solid. We will now investigate the effect of inhomogeneity on the buildup of self-organized phases of the
BEC-cavity system, and answer the question of how the different phases shown in Fig.~\ref{phase_transition} will manifest themselves in the experiment.
In contrast to the situation with pure contact interactions, we find that the properties of the
BEC in the optical cavity are strongly influenced by
the trapping potential, due to cavity-mediated long-range
interactions which are self-consistently determined by the density distribution of
the whole system. Here we consider a $N_{\rm lat}=32\times32$ lattice with harmonic trap strength $V_{\rm trap}=0.003E_R$. All other parameters are chosen as in Fig.~\ref{phase_transition}.
\begin{figure}[h]
\vspace{-35pt}
\hspace{-5mm}
\includegraphics*[width=3.6in]{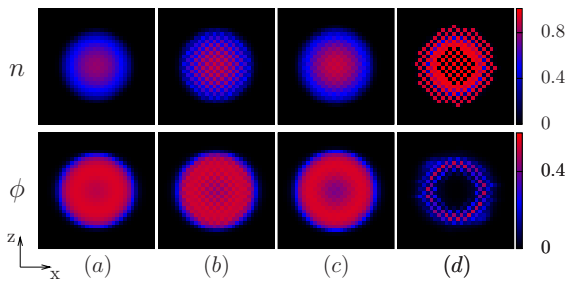}
\vspace{-50pt}
\caption{(Color online) Density distribution $n$ and superfluid
order parameter $\phi$ versus position on a square ($32\times 32$) lattice for
different atom numbers $N_{\rm tot}=139$, $167$, $184$, and $220$ in
panels (a), (b), (c), and (d), respectively. Other parameters are $\Delta_c=-500E_R$, $\kappa=300E_R$,
$U_0=-0.1E_R$, $V_p=15E_R$, with a harmonic trap
$V_{\rm trap}=0.003E_R$.}\label{density_trap}
\end{figure}
In Fig.~\ref{density_trap} we show the resulting density (upper
panels) and superfluid order parameter distributions (lower panels) in real space
for different total particle numbers. In general, the larger the
total particle number, the more photons are scattered into the
cavity mode, and thus the easier the system can form the
self-organized phase. We observe that at $N_{\rm tot}=139$, there is
almost no checkerboard phase region, as visible in panel (a). At $N_{\rm tot}=167$, the supersolid
phase can be clearly observed in the center of the trap, since with increasing $N_{\rm tot}$ the superfluid core expands at the trap
center and hence more photons are scattered into the cavity mode. From Fig.~\ref{phase_transition}, we
expect that the self-organized phase will disappear again when the
number of particles increases to a value at which a
Mott gap arises in the center of the trap, which is clearly
visible in panel (c) at $N_{\rm tot}=184$. After further increase of the particle number to
$N_{\rm tot}=220$, the checkerboard order reappears again. Moreover,
we observe that a checkerboard solid core with average filling $n=0.5$ builds up, indicating that the interplay between the trap inhomogeneity and cavity-mediated long-range interaction can give rise to new phases. Observation of these different phases is possible by using single-site addressing techniques in an optical lattice based on optical or electron microscopy \cite{M. Greiner_2009, I. Bloch_2010,single_site}.

In conclusion, we have investigated self-organized phases (supersolid and checkerboard solid) of both homogeneous and trapped
ultracold Bose gases coupled to a high-finesse optical cavity. We have found that these phases are robust
against strong onsite interactions at zero temperature, where the self-organization phase
transition is solely driven by quantum fluctuations. We observe that thermal fluctuations can enhance the buildup of self-organized phases at finite but low temperature.
In the presence of an external harmonic trap, the coexistence of
superfluid, Mott-insulating, supersolid and checkerboard solid domains
is observed. We find the buildup of these self-organized phases to be
strongly influenced by an external
trap, due to the density dependence of scattering between pump laser
and cavity mode by atoms in the cavity. Self-organized phases can be
detected by combining time-of-flight measurements and the detection of photons
leaking from the cavity \cite{T. Esslinger_2010}, while the
coexistence of different phases in the presence of an external trap
could be directly observed by quantum gas microscopy with single-site resolution
\cite{single_site, M. Greiner_2009, I. Bloch_2010}.

We acknowledge useful discussions with A. Hemmerich and R. Mottl. This work was supported by the China Scholarship
Fund (Y.L.), and by the Deutsche Forschungsgemeinschaft (DFG) via SFB-TR 49 and
the DIP project HO 2407/5-1. W.H. acknowledges the hospitality of KITP Santa Barbara, where
this research was supported in part by the National Science Foundation under Grant No. PHY05-25915.

\FloatBarrier


\begin{references}
%\bibliography{Literature.bib}
\bibitem{M. Lewenstein_2008} J. Larson {\it et al.}, New J. Phys. {\bf 10}, 045002 (2008).
\bibitem{H. Ritsch_2002} P. Domokos {\it et al.}, Phys. Rev. Lett. {\bf 89}, 253003 (2002).
\bibitem{A.T. Black_2003} A. T. Black {\it et al.}, Phys. Rev. Lett. {\bf 91}, 203001 (2003).
\bibitem{J. Reichel_2007_1} P. Treutlein {\it et al.}, Phys. Rev. Lett. {\bf 99}, 140403
(2007).
\bibitem{J. Reichel_2007_2} Y. Colombe {\it et al.}, Nature (London) {\bf 450}, 272 (2007).
\bibitem{P.W. Courteille_2007} S. Slama {\it et al.}, Phys. Rev. A {\bf 75}, 063620 (2007).
\bibitem{T. Esslinger_2006_2009_1} T. Bourdel {\it et al.}, Phys. Rev. A {\bf 73}, 043602
(2006); F. Brennecke {\it et al.}, Nature (London) {\bf 450}, 268 (2007); Science {\bf
322}, 235 (2008); S. Ritter {\it et al.}, Applied Phys. B {\bf 95}, 213 (2009); A. \"Ottl {\it et al.}, Rev. Sci. Instrum. {\bf 77}, 063118 (2006).
\bibitem{T. Esslinger_2010} K. Baumann {\it et al.}, Nature (London) {\bf 464}, 1301 (2010); K. Baumann {\it et al.}, Phys. Rev. Lett. {\bf 107}, 140402 (2011).
\bibitem{C. Maschler_2005} C. Maschler {\it et al.}, Phys. Rev. Lett. {\bf 95}, 260401 (2005).
\bibitem{J. Larson_2008} J. Larson {\it et al.}, Phys. Rev. Lett {\bf 100}, 050401 (2008).
%\bibitem{H. Ritsch_2008} C. Maschler {\it et al.}, Eur. Phys. J. D {\bf 46}, 545 (2008).
%\bibitem{H. Ritsch_2007} A. Vukics {\it et al.}, New. J. Phys. {\bf 9}, 255 (2007).
\bibitem{B. D. Simons_2009} M. J. Bhaseen {\it et al.}, Phys. Rev. Lett. {\bf 102}, 135301 (2009).
\bibitem{G. Morigi_2010} S. Fern\'andez-Vidal {\it et al.}, Phys Rev. A {\bf 81}, 043407 (2010).
\bibitem{H. Ritsch_2004} J. K. Asb\'oth {\it et al.}, Phys. Rev. A {\bf 70}, 013414 (2004).
\bibitem{P. Domokos_2007} J. K. Asb\'oth {\it et al.}, Phys. Rev. Lett. {\bf 98}, 203008 (2007).
\bibitem{T. Esslinger_2012}  R. Mottl {\it et al.}, Science {\bf 336}, 1570 (2012).
\bibitem{RBDMFT} Y.-Q. Li {\it et al.}, Phys. Rev. B {\bf 84}, 144411 (2011); Phys. Rev. A {\bf 85}, 023624 (2012).
\bibitem{third_direction} Here we choose the tight confinement $V_y=30E_R$.
\bibitem{P. Domokos_2008} D. Nagy {\it et al.}, Eur. Phys. J. D {\bf48}, 127 (2008).
\bibitem{W. Werger_2003} W. Zwerger, Journal of Optics B {\bf 5}, 9 (2003).
\bibitem{BDMFT_2009} A. Georges {\it et al.}, Rev. Mod. Phys. {\bf 68}, 13 (1996); K. Byczuk {\it et al.}, Phys. Rev. B {\bf 77}, 235106 (2008); A. Hubener {\it et al.}, {\it ibid.} {\bf 80}, 245109 (2009); W. Hu {\it et al.}, {\it ibid.} {\bf 80}, 245110 (2009); P. Anders {\it et al.}, Phys. Rev. Lett. {\bf 105}, 096402 (2010).
\bibitem{H. Ritsch_phase} D. Nagy {\it et al.}, Eur. Lett. {\bf 74}, 254 (2006).
\bibitem{Boninsegni} M. Boninsegni {\it et al.}, Phys. Rev. Lett. {\bf 95}, 237204 (2005).
\bibitem{finite-2D} N. Prokof'ev, Phys. Rev. A {\bf 66}, 043608 (2002); Z. Hadzibabic {\it et al.}, Nature (London) {\bf 441}, 1118 (2006); Z. Hadzibabic {\it et al.}, Riv. Nuovo Cimento {\bf 34}, 389 (2011).
\bibitem{W. Hofstetter_2010} U. Bissbort {\it et al.}, Phys. Rev. Lett. {\bf 106}, 205303 (2011).
\bibitem{I. bloch_2002} M. Greiner {\it et al.}, Nature (London) {\bf 415}, 39 (2002).
%\bibitem{M. Rafael} Private communication with M. Rafael.
%\bibitem{M. Lewenstein_2009} J. Larson and  M. Lewenstein, New J. Phys. {\bf 11}, 063027 (2009).
%\bibitem{P. Domokos_2010} D. Nagy {\it et al.}, Phys. Rev. Lett. {\bf 104}, 130401 (2010).
%\bibitem{P.M. Goldbart_2009} S. Gopalakrishnan {\it et al.}, Nature Phys. {\bf 5}, 845 (2009).
%single_site
\bibitem{M. Greiner_2009} W. S. Bakr {\it et al.}, Nature (London) {\bf 462}, 74 (2009).
\bibitem{I. Bloch_2010} J. F. Sherson {\it et al.}, Nature (London) {\bf 467}, 68 (2010).
\bibitem{single_site} T. Gericke {\it et al.}, Nature Phys. {\bf 4}, 949 (2009).
\end{references}
\end{document}